\documentclass[%
 reprint,
 amsmath, assume,
 aps,
]{revtex4-2}

\usepackage[colorlinks,citecolor=blue,linkcolor=magenta,urlcolor=blue]{hyperref}  

\usepackage{braket}
\usepackage{xspace}
\usepackage{tikz,quantikz}
\usepackage{orcidlink}
\usepackage{comment}
\usetikzlibrary{external}
\usepackage{amssymb}
\usepackage[T1]{fontenc}

\usepackage[ruled,vlined]{algorithm2e}
\usepackage[capitalise]{cleveref}	
\usepackage{import}
\usepackage{transparent}
\usepackage[normalem]{ulem} 

\usepackage[mode=build]{standalone}

\usepackage{graphicx}
\usepackage{dcolumn}
\usepackage{bm}
\usepackage{physics}	
\usepackage{dsfont}

\usepackage{multirow}

\crefname{algorithm}{Algorithm}{Algorithms}
\crefname{appendix}{Appendix}{Appendices}
\crefname{section}{Section}{Sections}

\begin{document}

\preprint{APS/123-QED}
\title{A resource-centric, task-based approach to quantum network control}
\author{ Alexander Pirker$^{1}$, Belén Muñoz$^{2}$ and Wolfgang Dür\orcidlink{0000-0002-0234-7425}$^1$}

 \affiliation{$^1$ Universität Innsbruck, Institut für Theoretische Physik, Technikerstraße 21a, 6020 Innsbruck, Austria}
 \affiliation{$^{2}$ Quantum Network Design GmbH, Technikerstraße 21a, 6020 Innsbruck, Austria}

\date{\today}

\begin{abstract}
Quantum networks exhibit fundamental differences from their classical counterparts. These differences necessitate novel principles when organizing, managing, and operating them. Here we propose an unconventional approach to organize and manage the operations of quantum network devices. Instead of a hierarchical scheme using layers, like in classical networks and present quantum network stack models, we propose a resource-centric task-based scheme. In this scheme, quantum applications pose objectives, initiated by a node, to a quantum network, such as sharing an entangled state or sending a qubit along a path. The quantum network node initiating the objective consequently derives a distributed workflow, referred to as saga, comprising numerous tasks operating on resources, which completes the objective. We identify three different kinds of resources with their own and independent topology, namely classical messaging, quantum channels and entanglement. Sagas can either be centrally orchestrated or performed in choreography by the network nodes. The tasks of a saga originate from and operate on resources of the network, such as quantum channels or entanglement, and they not only comprise operations and measurements, but potentially also include other tasks or even entire protocols, such as sending a qubit,  distributing entanglement or performing entanglement purification steps.
\end{abstract}

\maketitle

\section{\label{sec:intro}Introduction}

Quantum networks hold the promise to revolutionize information technologies and to play a key role in the success of current and future quantum technologies. In fact, connecting quantum devices serves as a crucial enabler to many different applications. For example, when coupling quantum computers, often referred to also as distributed quantum computing \cite{Beals20120686,CiracDistributed,CALEFFI2024110672,Cacciapuoti18,Caleffi18a}, the available state space for solving problems grows exponentially instead of linearly, as for classical computers. Not only quantum computers benefit from networks, also sensors exhibit higher precision when connecting them by entanglement \cite{Eldredge16,Sekatski2020,Hamann_2024,Hamann_2025}. Quantum networks offer many more interesting applications, especially in the field of security, such as key distribution, both bipartite (QKD) \cite{RennerPhd,LoQKD,ZhaoQKD,GottesmanQKD} or multipartite (conference keys) \cite{Chen07} protocols, for cryptographic protocols or secret sharing \cite{Hillery99,Markham08,Bell14}.

Unlike classical networks, no standard model or reference architecture, similar to the OSI model for computer networks \cite{Zimmermann88}, emerged yet for quantum networks. This has several reasons. First, of all, quantum networks correspond to a rather young field of research. Even though the connection of two ions over a network \cite{Krutyanskiy2023} or more nodes in a quantum network via optics \cite{Hermans2022,Main2025} have been experimentally demonstrated, practically speaking quantum networks are still far from being feasible. Building quantum networks poses several challenges, including short memory lifetimes, noise in quantum channels, probabilistic processes but also their organization and protocols. Especially the organization and protocols of a quantum network emerge as a complicated topic, since unlike classical networks, the utility of a quantum network goes beyond point-to-point transmissions of information. In particular, quantum networks allow nodes to share entanglement, a correlation between quantum systems not available in classical systems. Entanglement lies at the heart of countless quantum applications. Entanglement should not only be seen as a resource for applications, but also as resource at the protocol level and as part of the organization of quantum networks itself \cite{Cuquet2012,Pirker_2018,Pirker_2019,Meignant_2019}. Entanglement may serve to generate states from it, but also to implement remote gates via gate-teleportation or measurement-based quantum computing using the Jamiolkowksi isomorphism \cite{Jamiol}. Therefore, for a quantum network it is desirable to share entanglement between quantum network nodes. To distribute entanglement, but also to combat noise in channels efficiently, protocol and architectural efficiency of a quantum network is of utmost relevance \cite{Kolar2022,Cacciapuoti2024_1,Zhan2025}.

At the moment, three different models to organize a quantum network, referred to as stack models, have been proposed \cite{VanMeterStack1,VanMeterQRNA,WehnerStack1,WehnerStack2,WehnerStack3,Pirker_2019}. All of the models apply, similar to the OSI model, a hierarchical approach to tackle the complexity in a quantum network. Two of the models strive to distribute bipartite entanglement in terms of Bell-states \cite{VanMeterStack1,WehnerStack1,WehnerStack2,WehnerStack3} whereas the other one \cite{Pirker_2019} aims at distributing multipartite entanglement instead. The stack models introduce a variety of different layers, including layers to distribute entanglement, to generate long-distance entanglement or to establish network boundaries. However, all three models fall short from reflecting the resources and the dynamic nature of quantum networks and their tasks. In contrast to classical networks, quantum networks not only comprise channels but also entanglement as a resource. Considering entanglement in a network as a resource to accomplish tasks opens completely new possibilities. Furthermore, when acknowledging entanglement as a fundamental resource in a network, it becomes clear that any hierarchical layering in a quantum network control framework will introduce unnecessary complexities and difficulties, simply due to the rather dynamic and fragile nature of entanglement. Frankly, entanglement can exist across network boundaries, may appear as part of protocols over any kind of distance and therefore span multiple organizational layers. For example, establishing a long-distance Bell state from several short-distance Bell-states corresponds to a nested process in repeater networks \cite{1GRepeater}, potentially connecting two far distant networks through several intermediate networks. Entanglement may also change on demand, either within a network but also across network boundaries, thereby potentially establishing new network boundaries in terms of entangled states on demand. Lastly, also the task execution time in a network corresponds to one of the key performance indicators, due to the limited lifetime of qubits in memory. Hence, any unnecessary complexity in a control framework for quantum networks will lead to a degradation of states kept in memory. A hierarchical organization of entanglement therefore appears difficult due to very dynamic nature of combining/merging entanglement it into larger, but also smaller, states.

In this work we introduce a task-based approach as quantum network control framework. Instead of having a strict layer model, we pursue a resource-centric task execution model. Quantum applications pose objectives, like generating a GHZ state between nodes, to the quantum network, initiated by a quantum network node. The initiating node consequently derives a saga comprised of distributed tasks, involving also other nodes of the network, to achieve the objective. We lend the term saga from modern microservices architectures \cite{microservices}. In microservices architectures a saga corresponds to a distributed workflow comprised of tasks executed by different services \cite{saga}. For our quantum network control framework, a saga corresponds to a set of distributed tasks or protocols carried out by nodes of the quantum network. Tasks associate with a resource, and may correspond to an entire protocol. For example, consider the task of executing a Midpoint-source protocol (MSP)\cite{Jones_2016}. In the Midpoint-source protocol, an entanglement distribution task, two end nodes connect to a central node in terms of a quantum channel. The central node generates a Bell-state locally and sends one qubit as photons to each end node. The end nodes convert the photonic/flying qubit into a matter qubit and keep the state. This results in a Bell-state between the two end nodes. Alternatively, as referred to as Meet-in-the-Middle protocol in \cite{Jones_2016} or in the community as Midpoint protocol, one could prepare Bell-states at the end nodes, send half of them to a central node which ultimately performs a Bell-State-Measurement (BSM) to distribute an entangled state among the end nodes. Another task corresponds to sending a qubit, a channel tasks involving only two nodes. Tasks can run sequentially, or in parallel, and they utilize resources. We identify in total three different kinds of resources in a quantum network, namely classical messaging, (quantum) channels and entanglement. Some of the tasks, such as entanglement distribution tasks, generate new (entanglement) resources, which mandates dynamic updates of the available resources. To derive a concrete saga achieving an objective, a quantum network node inspects the current resources of the network, and based on metrics on these resources the network node identifies a series of tasks that achieve the objective, resulting in a saga. We note that a saga in itself corresponds to some sort of protocol, which will also mandate classical messaging as part of its execution. To facilitate the urgency of objectives, we further introduce priorities to objectives and sagas. These values allow for the consideration of priorities in the derivation of sagas. We note that this work focusses on laying out the theoretical foundation of the quantum network control plane.

The paper is organized as follows. In section \ref{sec:background} we give a brief review of classical network models, and state-of-the-art quantum network stacks and quantum network control frameworks. Section \ref{sec:model} introduces our new framework. There we discuss resources in more depth and go into details regarding sagas. We give an outlook and conclude our work in section \ref{sec:outlook}.

We note that in this paper we consider qubits, however the entire control framework is directly applicable also to higher dimensional quantum states.

\section{Background}\label{sec:background}

\subsection{Classical network protocol stack}\label{sec:sub:classical}

\begin{figure}
    \centering
    \includegraphics[width=\columnwidth]{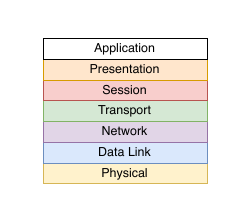}
    \caption{The standard OSI model for classical computer networks. It comprises in total seven, hierarchically organized, layers. Data passes from layer to layer in terms of a packet. Each layer adds additional information specific for its layer to the packet, referred to as header. Each layer only operates on the header information belonging to its layer.}
    \label{fig:osi}
\end{figure}

Since decades the classical internet uses the OSI model for computer networks \cite{Zimmermann88}. Since its development in the late 1970s, the model became the de-facto standard for nearly all modern computer networks, including the internet. The model is rather simple, as it breaks down complex tasks within a network into multiple hierachical layers. Each of the layers has a dedicated responsibility, such as facilitating flow control, or ensuring data transmission on a medium. Information, also referred to as packets, passes from top layers to bottom layers, and each layer implements a particular protocol operating exclusively on this layer. Each protocol adds additional information to packets, such as IP addresses for the network layer \cite{ip} or port numbers for the transport layer \cite{tcp,udp}. The model is also depicted in Fig. \ref{fig:osi}.

The top-most layer is the application layer, corresponding to the application using the network. Such an application could be a browser, but also an email client. Whenever an application sends information through a network, the information gets wrapped in a packet, and passed down to the presentation layer and the session layer. From there, the session layer forwards the packet to the transport layer to implement transport control mechanisms for the packet transmission. For that purpose the transport layer adds additional headers to the existing packet, such as control flags and port numbers. After that, the packet is passed on to the network layer, which is responsible for the logical decomposition of the network. It implements mechanisms to address network devices and adds the corresponding addresses to the packet as further additional header. The data link layer also adds some information to the packet such as MAC address, and ultimately passes it to the physical medium. At each intermediate node and especially also the end node, the packet gets unpacked (and potentially repacked) in the opposite order of layers.

As we can see, the model implements a strict hierarchical control framework for classical networks, in which layers implement protocols that append additional information to an existing packet in order to facilitate transfers. At its heart, and of fundamental importance, is that devices can freely copy information without destroying it. Furthermore, even though real-time applications demand high throughput and performance, the whole processing pipeline of packets up to higher layers is still very fast in contrast to memory lifetimes. 

We point out those features, as for quantum networks the situation is fundamentally different.

\subsection{Quantum networks and their differences to classical networks}

Before discussing the state-of-the-art quantum network control frameworks (and their limitations), we briefly recap the main differences between classical networks and quantum networks. 

In classical computer networks the sole responsibility of the network is to transport bits between two nodes. For that purpose the nodes of networks utilize the hierarchical OSI reference architecture model, as described in section \ref{sec:sub:classical}. Following the OSI model one categorizes the network devices by their respective operating layer, like for example routers (network layer), switches (data link layer), hubs (physical layer) or computer end nodes (application layer). Depending on the layer a network device operates on, the device unwraps packets until its operating layer, inspects the header of the layer, and takes a decision how to handle and forward the packet based on the information within the header. Therefore, the sole purpose of a network device corresponds to processing a packet on its operating layer and forwarding it to the next network device until the destination node is reached. All this classical processing introduces waiting times, referred to as latency. The time until a packet reaches an end node depends heavily on the traffic load of the network. The processing approach in general of a network device corresponds to a store-inspect-restore scheme, where a classical device keeps the packet in memory, reads it, and restores it when forwarding it.

Quantum networks, at their core, transmit qubits between remote nodes through noisy channels. When sending a qubit through a noisy channel, for example through fiber or free-space links, the traveling qubit will interact with the environment, resulting in noise on the qubit. This noise essentially destroys the information held by the qubit. Unlike classical networks, where network devices read information, restore it, and retransmit it, quantum network nodes cannot copy or clone information (due to the no-cloning theorem \cite{NoCloning}). This fundamentally limits the distance of direct transmission of unprotected qubits in a network. And this fundamental difference from classical networks makes quantum networks so hard to operate, as the classical approach of store-inspect-restore is not applicable to quantum networks. Specifically, a quantum network device cannot simply inspect and copy the information held by a qubit in general, because any measurement will destroy the information stored in the qubit. To overcome this fundamental limitation, several schemes and protocols were developed, referred to as quantum repeaters  \cite{1GRepeater,ZwergerDFS,Jiang2009,DurRepeater,Azuma2015,VanMeter2014book}. Many of these schemes utilize entanglement at their heart to overcome large distances, and sending qubits can then be implemented via quantum teleportation \cite{BennettTele}. 

But not only the transmission of qubits is fundamentally different from classical networks. Recall that the only responsibility of a classical network corresponds to the transport of data between two nodes of a network. In contrast to its classical counterpart, quantum networks offer much richer functionality and services than classical networks, as nodes running on a quantum network may wish to establish entanglement instead of transmitting qubits only. Entangled systems exhibit  correlations not accessible in classical system. This correlation is not limited to two parties only, but also exist between multiple systems, referred to as multipartite entanglement. Entanglement is the key ingredient to many applications including distributed sensors \cite{Eldredge16,Sekatski2020,Hamann_2024,Hamann_2025}, secret sharing \cite{Hillery99,Markham08,Bell14} but also distributed quantum computing \cite{Beals20120686,CiracDistributed,CALEFFI2024110672,Cacciapuoti18,Caleffi18a}. Such applications are believed to either outperform existing applications, or even lack a classical counterpart entirely. Hence, the services a quantum network offers go far beyond classical networks and make quantum networks entirely different from classical networks. We conclude therefore that one of the ultimate goals of a quantum network corresponds to sharing entanglement among its end nodes. 

In a quantum network nodes carry out classical processing tasks but also quantum information processing tasks in parallel, for example in terms of unitaries or measurements, to implement tasks. For instance, for entanglement swapping, a quantum network node applies a Bell-state measurement, and sends the outcome to another node for further processing. Depending on this outcome, the node will apply potentially a correction operator. The timing of applying such operations is vital to the correct functionality of a quantum network, and do not exist in classical networks.

Lastly, and most importantly, memories in classical networks and quantum networks operate at entirely different time scales when it comes to their lifetimes. While  information in classical devices may reside in memory for a very long time without suffering noticeable noise, the lifetimes of qubits in quantum memories does not exceed a couple of seconds. This major difference implies that any framework to operate a quantum network should be streamlined to performance and efficiency, and minimize any overhead. 

\subsection{State-of-the-art quantum network control frameworks and their problems}

Up to present day, mainly three different stack models for quantum networks have been pursued in the literature \cite{VanMeterStack1,VanMeterQRNA,WehnerStack1,WehnerStack2,WehnerStack3,Pirker_2019}. The stack models have in common that all of them pursue a similar approach as in classical networks. In fact, all of them break down the complexity of a quantum network and its control plane into hierarchical layers. In the following we discuss the individual models briefly. An excellent discussion and in depth comparison about the three different models can be found in \cite{Illiano2022}.

The stack model of \cite{VanMeterStack1} uses at its lowest layer a physical entanglement (PE) layer. This layer is responsible for generating entanglement between two neighboring nodes. The next layer, namely the entanglement control layer, determines whether entanglement attempts where successful or not. On top of these two layers reside purification control layers and entanglement swapping control layers. The model nests these layers to support the generation of long-distance entanglement between two nodes in a quantum network. Ultimately, at the highest layer, the model comprises an application layer, corresponding to the application running in the network requiring entanglement. We note that the main purpose of this model is to distribute bipartite entangled states between nodes, and thereby focuses on end-to-end or point-to-point entanglement in terms of Bell-states. This basic model was further developed  into the Quantum-Recursive-Network-Architecture (QRNA) \cite{VanMeterQRNA} years after its first publication, again focusing on the generation of Bell-states.

Also the model of \cite{WehnerStack1,WehnerStack2,WehnerStack3} aims at generating bipartite entanglement by utilizing several hierarchical layers. In contrast to the model of \cite{VanMeterStack1}, it introduces a hardware abstraction layer, and the notion of a link layer. The goal of the lowest layer, namely the physical layer, is to generate entanglement between two nodes in defined timeslots using the Midpoint-Heralding-Protocol (MHP). In the MHP, two end nodes connect to a central node in terms of channels. The end nodes locally prepare Bell-states, and send one of the qubits each to the central node. The central node consequently performs a heralded Bell-state measurement, which, if successful, generates entanglement between the end nodes. The physical layer actually polls the link layer if entanglement should be established or not, and takes all necessary actions. Whether an entanglement attempt should be done or not is actually decided at higher layers, such as the network layer. In fact, the network layer is responsible for establishing long-distance entanglement, and it uses the functions defined in the link layer, such as entanglement swapping amongst others, to fulfill this task. The transport layer operates on top of the network layer, and is responsible for transmitting qubits using quantum teleporation. Hence, also here we note that the ultimate goal of the model is to generate Bell-states between two nodes of the quantum network.

In contrast to the other two models, the model of \cite{Pirker_2019} aims at distributing multipartite entanglement (in terms of graph states \cite{heinEntanglementGraphstates}) instead of bipartite entanglement. Here, the stack comprises several hierarchical layers. The lowest layer, namely the physical layer, is responsible for sending qubits between nodes and connecting quantum network nodes. The connectivity layer operates on top of the physical layer, and is responsible for establishing long-distance entanglement in terms of multipartite states. The next layer, namely the link layer defines the boundary of a network in terms of multipartite states, and aims at generating arbitrary graph states between network nodes. Ultimately, the network layer connects different quantum networks again via multipartite entangled quantum states. 

Other models aiding the control within a quantum network have been recently published in \cite{wehnerscheduling,LopezStrata,qnodeos,Schon2024,Chung2022,vanmeterQuInternetArchitecture2022}. The work of \cite{wehnerscheduling} uses a centralized scheduling mechanism which executes tasks in a network in time slots. Having a centralized scheduling for entanglement generation simplifies the network operation, however, it also makes the network fragile to failures or overloading of the scheduler, as the scheduler becomes the bottleneck of the network. In contrast, the framework of \cite{LopezStrata} proposes a framework composed of so-called "strata". In total, the work identifies three different strata, namely the service strata, the connectivity strata and the compute strata. The service strata contains the functions offered to external applications, whereas the connectivity strata is responsible for functions that focus on the transfer of data between communication endpoints. Ultimately, the compute strata addresses the control, management and resources planes for computing services and aspects. The work of \cite{qnodeos} demonstrates a quantum network operating system. This work presents a viable architecture of an operating system running on a quantum network node, aiming at generating entanglement between end-nodes. It is important to note that it actually utilizes the network stack of \cite{WehnerStack1,WehnerStack2,WehnerStack3} rather than proposing one. Similarly, the work of \cite{Schon2024} also demonstrates an experiment using a centralized quantum network controller orchestrating several kinds of network nodes, such as quantum end nodes, quantum repeaters, Bell-state measurement nodes and quantum channels. Also there, the quantum network server coordinates all activities in the network by using a message bus implementation and broker service. The work of \cite{Chung2022} distinguishes two layers for the quantum network control plane, namely a device control functions layer and a network control function layer. Each of these layers contains several network functions such as time synchronization or optical path routing and wavelength assignment. These functions ultimately serve as a foundation to implement a plane offering services, such as entanglement distribution or quantum teleportation. We note that the focus of our work here is on a general quantum network control framework, going beyond the generation of entanglement between two end-nodes and offering elementary services. The goal is to formulate a framework to compose functions of/and services in a quantum network into larger distributed, yet coordinated, workflows to complete an objective, potentially involving several nodes, in a modular manner. In contrast to the work of \cite{vanmeterQuInternetArchitecture2022}, which proposes static RuleSets to handle and manage entanglement generation we target the dynamic composition of tasks into distributed workflows. The community also progressed on languages to instantiate and program network models recently. The work in \cite{Kozlowski2024} introduces a language to actually implement parts of quantum network control rather than proposing a control framework itself, and was used to implement parts of the stack of \cite{WehnerStack1,WehnerStack2,WehnerStack3}. The work of \cite{Dahlberg_netqasm} proposes also a programming language on top of OpenQASM to implement quantum network application, and builds on top of a quantum network stack model such as \cite{WehnerStack1,WehnerStack2,WehnerStack3}.

All of these models and proposals fall short by reflecting and taking into account for the various types of resources in quantum networks. Unlike classical networks where the only resource corresponds to classical messaging, quantum networks utilize several different kinds resources to fulfill tasks. These resources corresponds to (quantum) channels, pre-shared entanglement and classical messaging. In the remainder of this work we use the term channel instead of quantum channel whenever the context is clear. The model we propose in section \ref{sec:model} centers around resources rather than hierarchies. We note that just identifying and organizing a quantum network around resources constitutes not only a fundamental difference to the classical network stack, but also a major shift in paradigm with regards to current quantum network control frameworks and organizational models. As stated above, the majority of current quantum network models introduce and utilize layers and hierarchies, categorizing a particular resource into one of these layers. However, here we rather strive for not imposing any layer or hierarchy on a resource.

Network nodes use channels as a resource to send qubits. In contrast, entanglement may serve as a resource for implementing remote gates via gate-teleportation or performing measurement-based quantum computing using the Jamiolkowski isomorphism. Entanglement transforming tasks also include for example merging two graph states, establishing (long-distance) Bell-states or performing entanglement purification. Therefore, we conclude it is natural for a quantum network control framework to distinguish between these different resources.

We also find that in quantum networks there may exist many different ways how to achieve a goal (or how we call it later, an objective), unlike in classical networks where the only goal corresponds to sending bits between two nodes. For example, sending a qubit can be either done by transmitting it directly through a channel but also by consuming a pre-shared Bell-state by using quantum teleportation. Similarly, generating a Bell-state between two nodes having a central node in between can be done by either using a Midpoint-source protocol, by using a Midpoint(-Heralding-)protocol or by sending one half of a Bell-state through a channel till the end node. Which task suits best in certain situations depends highly on several environmental factors such as pre-shared entanglement, timings and channel quality, among others. This highlights the necessity to decouple objectives from a fixed implementation of how to achieve the objective. In the remainder of this paper we will frequently use the term Midpoint protocol, which either refers to the Midpoint-Heralding protocol or its counter part without the heralding option (i.e. Bell-state measurement without heralding).

Furthermore, all current frameworks aim at squeezing entanglement into a hierarchical order. In contrast to this, entanglement rather appears as a recurring element at any hierarchical level of a network. For example, establishing a Bell-state between two neighboring nodes by performing a Bell-state measurement in a middle node (Midpoint-protocol) corresponds to the very same task, in terms of quantum information processing operations, as establishing a long-distance Bell-state via entanglement swapping on already established Bell-states. The resulting resource is, in both cases, a new Bell-state, just spanning either a short distance in a network or a long distance. Furthermore, as we infer from this trivial example, entanglement is a transient resource, and may change and evolve over time. For example, in an entanglement-based quantum network, where the topology of the entanglement comprises the network's boundaries \cite{Pirker_2018,Pirker_2019,Mazza2025}, it is straightforward to extend the boundary of such a network by fusing states from different networks together. This changes the (entanglement) topology of a quantum network. Also within the very same network even, entanglement may be manipulated by local operations, like for example merging two states into one or measuring parts of a state. From this we deduce that any hierarchical organization of responsibilities in quantum networks concerning entanglement, as up to present day, simply falls short from reflecting the very dynamic nature of entanglement and its transformations. 

Ultimately, unlike in classical networks, quantum information decoheres very fast. From this it is clear that the execution of tasks and their organization is vital. In a layered architecture, each layer introduces additional performance overhead for its processing logic, and therefore contributes to the degradation of quantum states kept in memory. This highlights that entanglement is a rather dynamical, sometimes even short-lived resource, as it degrades over time. If entanglement is not taken care of actively by applying entanglement purification on a regular basis, it will decohere and become useless rather fast. Hence, a more direct and efficient approach to solving tasks in a quantum network is of high relevance to ensure fast processing of quantum information. This implies that a quantum control framework should simplify and streamline structures and executions as much as possible to minimize the imposed control overhead.

\section{Resource-centric, task-based quantum network control framework}\label{sec:model}

\begin{figure}
    \includegraphics[width=\columnwidth]{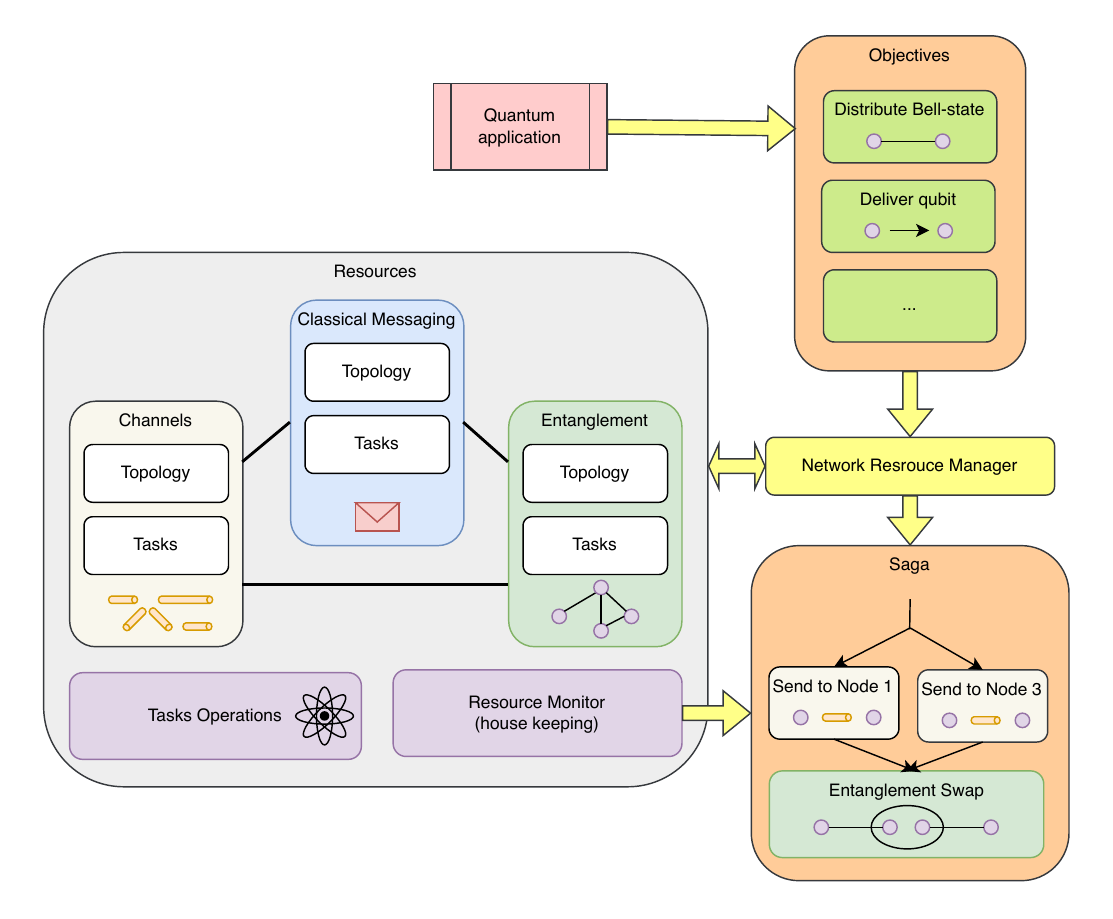}
    \caption{The figure depicts the proposed quantum network control framework. Quantum applications create objectives, like for example distributing a Bell-state. The network resource manager of a node uses its knowledge about the resources of the network node (channels, entanglement, classical messaging) to compute a saga, consisting of distributed tasks and protocols, to achieve the objective. The quantum network nodes consequently execute the saga to achieve the objective.}
    \label{fig:overallmodel}
\end{figure}

The model we propose is depicted in Fig. \ref{fig:overallmodel}. The execution starts with a quantum application running on a quantum network node demanding an objective from the network. An objective describes a certain goal, like for example sharing a Bell-state, or sending a qubit. The quantum network node consequently instructs the network resource manager, a component running on a node and managing the resources known to a quantum network node, to derive a series of tasks achieving the objective, referred to as a saga. The network resource manager composes such a saga from individual tasks of the resources of a quantum network. We note that a saga is a distributed workflow comprised of distributed tasks, executed by the nodes in the network. Tasks correspond to the elementary actions that a quantum network node or a set thereof can execute, such as sending a qubit through a channel or generating entanglement with a neighboring node using an entanglement distribution protocol. We discuss tasks in more depth in section \ref{sec:sub:sub:tasks}. Tasks can also be related to network function virtualization (NFV) \cite{nfv}, as tasks, as we use them here, correspond to individual building blocks to compose a distributed network workflow. We point out that, similar as in NFV, a goal may be achieved more than a single task which allows for the virtualization of how to achieve a goal. Furthermore, sagas relate to the term service function chaining (SFC) in classical networking, as a saga composes several tasks into a meaningful, chained, network workflow to achieve an objective. The resources of a quantum network node correspond to channels, entanglement and classical messaging, and each of them contains a set of tasks available to the network resource manager(s). Resources also include elementary operations, such as unitaries or measurements. Tasks can run sequentially, or in parallel.

We observe that in order to compose a saga for an objective, the network resource manager requires information concerning the resources in the network, like for example the channel topology but also the entanglement topology in terms of pre-shared entangled states. We propose to distribute topological information regarding any changes to any of the topologies of a network by broadcasting messages to other nodes in the same network. This ensures a unified and holistic view on the resources of the quantum network for each quantum network node. In the remainder of this section we describe in more detail the individual model elements.

However, we point out that our entire quantum network control framework can also be implemented by centralizing the network resource manager and its topology views into a centralized network manager, similar in spirit to the entanglement defined controller in \cite{caleffi2025quantuminternetarchitectureunlocking} or the centralized scheduling in \cite{Schon2024} but going beyond their capabilities. The centralized network resource manager manages the topological views of the network but also derives sagas centrally upon requests for objectives of nodes. After composing a saga, the network resource manager distributes the saga to the network nodes that consequently execute it. The advantage of a centralized management component is that it minimizes the communication overhead and simplifies the composition of sagas. However, it also constitutes a single point of failure and may become a bottleneck when deriving sagas. 

\subsection{Objectives}

The ultimate purpose of a quantum network is to assist quantum applications. In many applications, nodes must exchange information or generate entanglement. However, a quantum application should not need to know about the topological details of a quantum network, and instead rather demand the network to achieve an objective. It is important to note that such an approach hides all details concerning the network and its technological stack from the quantum application.

An objective in our framework refers to a goal that the quantum network should achieve. However, it falls short from having a concrete implementation, as it solely describes the goal rather than the way how to achieve it. For example, sending a qubit from one node to another in a quantum network can either be done by directly sending the qubit through a channel but also by utilizing a shared Bell-state in terms of quantum teleportation. The decision which of these tasks is better suited to achieve the objective depends on many factors, such as the quality of the channel connecting the nodes or whether the nodes already share a Bell-state or any other kind of entangled state that could be converted into a Bell-state, such as GHZ state shared with other nodes.

Objectives may have several parameters such as fidelity or  priorities. For instance, two nodes may wish to establish a Bell-state of certain fidelity, and hence this fidelity corresponds to a parameter of the objective. A priority of an objective reflects the urgency to achieve it, which helps the network resource manager of a quantum network node to prioritize resource selection and saga execution.

\subsection{Resources}

A quantum network comprises resources of different kinds. We identify three elementary resource kinds in a quantum network, namely quantum channels, entanglement and classical messaging. We note that the three resources influence each other. For example, sending an entangled qubit through a channel will change the entanglement topology of the quantum network. Each resource kind comprises a topology, and a set of tasks a node can execute. We note that there may exist several other kinds of resources that our framework not covers yet, however, they should be seen as a starting point.

\subsubsection{Topology}

Each of the resource types has a topology. The topology of classical messaging captures the classical communication channels between the network nodes. 

The channel topology reflects the quantum channels connecting the quantum network nodes with each other. A channel is characterized by its type and the associated (noise) parameters, like for example distance, absorption, photon-loss etc. This information is vital when composing sagas, as it describes the amount and type of noise a qubit will be exposed to on its transmission.

The entanglement topology corresponds to all the entangled quantum states shared across network nodes. For the entanglement topology we observe that the kind of entanglement constitutes vital information due to the existence of different entanglement classes. For example, for three qubits there exist two entanglement classes that can't be transformed into each other \cite{WState}. For graph states \cite{heinEntanglementGraphstates,HeinPRA2004}, a special class of multipartite entangled quantum states, the entanglement correlations between qubits can be described by a classical graph $G=(V,E)$. The vertices correspond to the qubits, and the edges to the entanglement correlations. Operations on graph states, such as measurements or local complementation, can be depicted by graphical manipulation rules on the corresponding classical graph. For graph states it is not sufficient to solely keep track of the nodes which are entangled, as graph states have entanglement correlations corresponding to edges in a classical graph rather than just its vertices. Therefore, also here topological information is of utmost importance. And for general entangled states the situation is even more complicated, as there is no simple notion of topology like for example a classical graph. Not only the information concerning the structure/topology of an entangled state is important. Entangled states can be transformed on-the-fly via local operations and classical communication (LOCC). We note that such operations may change the entanglement structure of a state. For example, performing a Pauli Z measurement on a graph state will remove the measured qubit from the graph and remove all edges incident to the measured qubit. Or performing a local complementation on a vertex of a graph state will invert the subgraph induced by the neighborhood of the qubit subject to local complementation. From this we find that the entanglement structure of entangled states in a quantum network is highly dynamical.  Not only the structure is of relevance, also the fidelity of a state matters for applications, because states will degrade over time when kept in memory. Therefore, quantum network nodes must also estimate expected fidelities of entangled states after establishing them from time to time.

We emphasize that the channel topology of the network can, and probably will, be completely different from the entanglement topology of  the network. This implies also that there will be two different kinds of networks overlaying each other, namely a channel network and an entanglement network \cite{Pirker_2018,Pirker_2019,Meignant_2019,mazza2024,mazza25}. 

\begin{figure}[h!]
    \includegraphics[width=0.9\columnwidth]{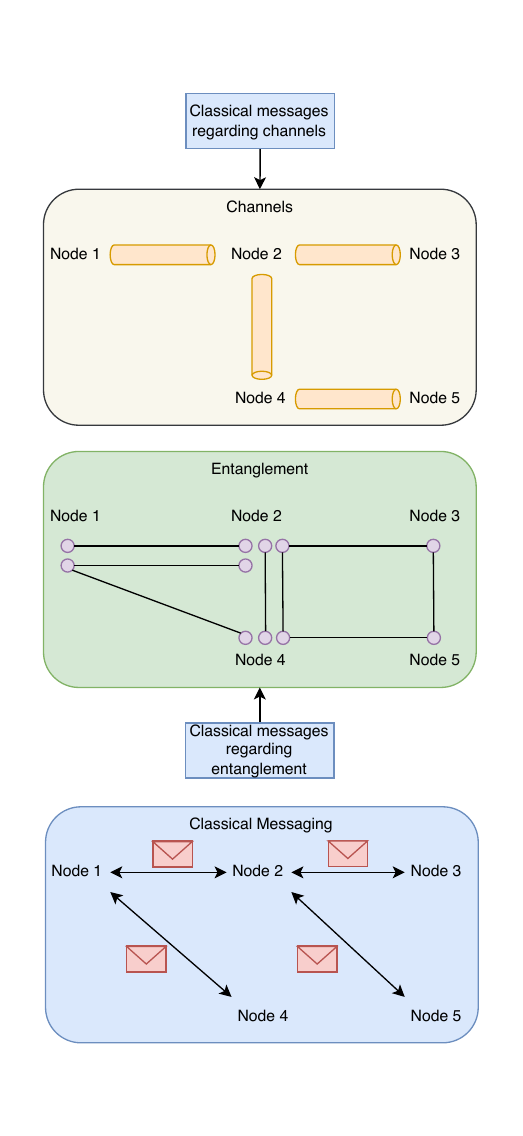}
    \caption{Every quantum network node has a global view on all the resources available in a quantum network. Classical messaging changes this view, for example when new entanglement was established, or channels become available.}
    \label{fig:topology}
\end{figure}

The topology of a resource corresponds to a vital parameter when composing a saga for an objective. We assume that all network nodes share the same global topological view on the network resources. For example, each quantum network node knows all the channels (and their noise models) within the quantum (channel) network. Similarly, all quantum network nodes also know about all entangled quantum states shared within the network, their structure, and their fidelities when initially generated. The situation is summarized in Fig. \ref{fig:topology}. 

To achieve such a global view, the quantum network nodes broadcast classical messages concerning topological changes through the entire network. This allows all other nodes to update their topological view accordingly. We note that this scenario also takes into account for dynamic events, such as the availability of a quantum channel to a satellite in certain time frames. We note that such an approach necessitates the segmentation of large-scale networks into smaller, more manageable, networks to achieve an efficient distribution of the necessary information for all nodes. In case of large networks it may take too long for the global view to converge in case all nodes share the same topological view on all the resources of the quantum network. However, this aligns naturally with other assumptions on networks, as they require precise timing and synchronization for many tasks such as entanglement distribution.

For entanglement resources it will be also crucial to implement locking strategies and update the list of entangled nodes on any changes. For example, two applications can't use the very same entanglement resources in two concurrent sagas. The network resource managers should send information concerning locks of entanglement for particular sagas to the other nodes prior to executing a saga. This prevents the accidental use of entanglement in multiple sagas. We outline here three different strategies to facilitate the locking of resources.

\begin{itemize}
    \item Decentralized Pessimistic Locking: When constructing a saga, the network resource manager of a node marks the resource locally in its topology view as "blocked, not confirmed", and sends a request to the nodes having parts of the entangled states. In case the entanglement is free, the parties of the entangled state mark the entanglement as "reserved" and respond to the saga initiating node with an acknowledgment message, and the nodes start an expiration timer. If the entanglement is already reserved, the nodes respond with a negative acknowledgment to the network resource manager initiating the saga. The network resource manager collects the acknowledgments of the lock, and continues with the construction of the saga. In case the saga composition completes successfully, the nodes consume the "reserved" states as part of the saga. Otherwise, the lock either expires due to the timer on reservation, or the network resource manager of the intiating node may also release the lock on those entangled states. This approach appears rather similar in spirit to the protocol described in \cite{Zhao2024}.
    \item Decentralized Optimistic Locking: Unlike the pessimistic locking, the network resource manager of a node may just simply go ahead and attempt to execute a saga without worrying about locking an entangled state at all. This saves time, but may result in failed sagas as nodes will compete without locks for a limited set of resources.
    \item Centralized Entanglement Lock Monitor: To minimize the communication between nodes, one alternative is to introduce a centralized entanglement monitor whose sole responsibility is to keep track of locks on entangled states. Similar as in the decentralized pessimistic approach, a network resource monitor queries the entanglement lock monitor if an entangled state is already reserved. If not, the network resource manager of a node reserves it for consumption in a saga. This scheme reduces the messaging overhead imposed by a decentralized locking strategy yet it still benefits from the distributed nature of saga derivation. It is similar in spirit to the entanglement defined controller discussed in \cite{caleffi2025quantuminternetarchitectureunlocking}, which is responsible for orchestrating the entanglement resources in a quantum network.
\end{itemize}

Ultimately we cover how to track entanglement in the quantum network. For that purpose we differentiate between "intermediate" entanglement that appears during the execution of a saga and the output entanglement of a saga. Intermediate entanglement doesn't have to be tracked as it solely appears as an intermediary result to achieve the objective of a saga. However, the output entanglement of a saga, if there is one, is of value to the network and hence requires tracking. For example, suppose that the resource monitor of quantum network is configured to keep a certain entanglement topology, i.e. a network state, alive. To generate such a network state, the resource monitors use sagas, and the network state corresponds to the output of those. This can be useful for fast responses to objectives, as the entanglement may serve as a short-cut to complete tasks faster. To keep track of such entangled states, the network nodes exchange classical message on establishing the state for the first time (Entanglement-Created-Message). Such a message contains an identifier for the state, the parties of the entangled state, the class of entanglement, the last accessed time, together with an estimate of its fidelity. Every time the entangled state changes, e.g. due to merging, the network nodes exchange an Entanglement-Updated-Message, communicating the changes among the state. When deriving a saga the network resource manager extrapolates the expected fidelity by taking into account for the fidelity, the last accessed time, and the current time together with the noise model of the storage(s). Therefore, to track entanglement, our framework requires an initial message communicating the generation of the entangled state and updates when modifying the topology of the entangled state.

\subsubsection{Tasks}\label{sec:sub:sub:tasks}

\begin{figure}[h!]
    \centering
    \includegraphics[width=\columnwidth]{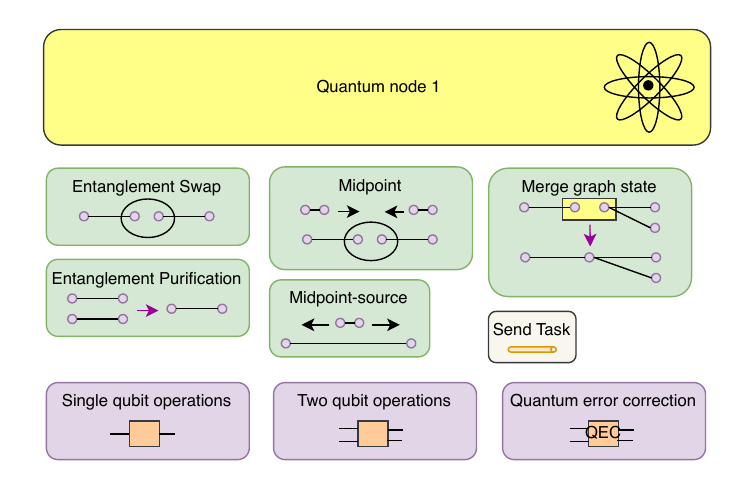}
    \caption{The figure depicts the tasks that quantum network node 1 offers. The node offers five entanglement tasks, like entanglement-swap or merging graph states. For channel resources it can send qubits. Lastly, node 1 can implement arbitrary operations (including also quantum error correction).}
    \label{fig:tasks_node1}
\end{figure}

In addition to topological information, each of the resources  contains tasks. Tasks form the most elementary executable parts of a quantum network. To execute a task, quantum network nodes classically communicate and carry out local, yet coordinated, quantum operations. Therefore, such tasks correspond to perfect candidates for formalizations and standardizations. A task for a resource specifies concrete steps to achieve a certain, yet elementary, objective. For example, for channels there exist one task for each pair of nodes connected through a quantum channel. The task corresponds to sending a qubit through such a channel. In contrast, for entanglement resources there exist numerous tasks. For instance, to distribute entanglement, a quantum network node offers tasks such as Midpoint or Midpoint-source protocol. Other tasks that manipulate entanglement include entanglement purification \cite{Bennett96,Bennett96_2,Deutsch96,Dur2007,Riera2021}, entanglement swapping by means of teleportation \cite{BennettTele}, merging graph states \cite{heinEntanglementGraphstates,Pirker_2018} or graph state fission \cite{miguelramiro2024graph}. Also entanglement itself can be seen as a task. For example, an entangled state can be used to perform a non-local gate via gate-teleportation or an entire circuit via measurement-based quantum computing using the Jamiolkowski state of the circuit. For such entanglement tasks, however, we point out that the availability of the task heavily depends on the fidelity of the entangled state, and that the executability of the task diminishes as soon as the fidelity of the state decreases too much. Fig. \ref{fig:tasks_node1} summarizes the situation for quantum network node 1.

\begin{figure}
    \includegraphics[width=\columnwidth]{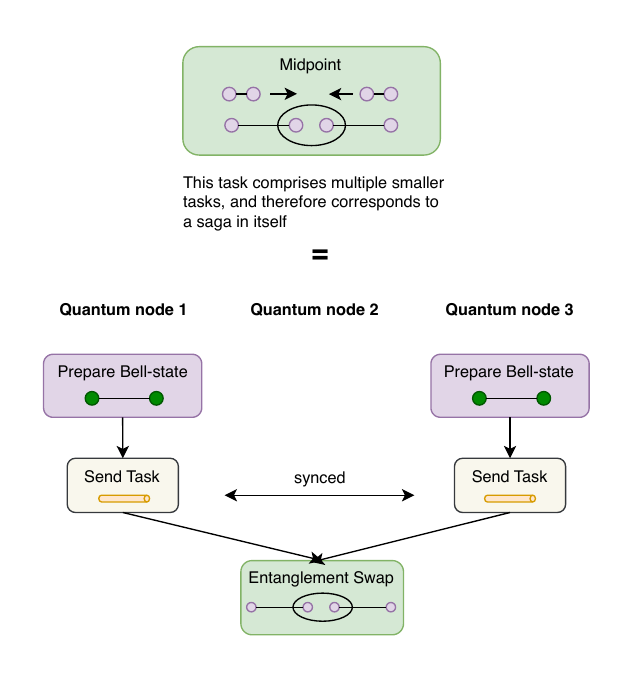}
    \caption{The Midpoint task/protocol can be decomposed of several other elementary tasks. In fact, it comprises two Bell-state preparation tasks followed by two synchronized send tasks and an entanglement swapping step in the central node.}
    \label{fig:task_decomposition}
\end{figure}

Tasks can itself again contain multiple subtasks to achieve their goal, see also Fig. \ref{fig:task_decomposition} as an example. For instance, the Midpoint task chains Bell-state preparation tasks together send qubit tasks and an entanglement swap task, and therefore makes use of tasks from multiple resources.

But resources also include elementary operations such as unitaries or measurements. These kinds of elementary operations correspond to a vital part in many tasks, and therefore must be available when composing tasks.

The resource monitor depicted in Fig. \ref{fig:overallmodel} maintains entanglement in case it was distributed in the past but shall be kept for future requests. For that purpose the resource monitor runs tasks to distribute additional entangled states in the network followed by entanglement purification steps consuming them on a regular basis. This increases the fidelity of stored entangled states shared in the network, and aids to fast response to emerging entanglement requests from the the network.

To conclude this section of tasks, we provide an non-exhaustive list of tasks for quantum networks, categorized by their resource kind. We emphasize that this list is not considered complete by any means, but should rather be seen as a starting point for future works.

\begin{itemize}
    \item \textbf{Channels}:
    \begin{itemize}
        \item \emph{Sending a qubit}: This task corresponds to sending a qubit through a channel.
    \end{itemize}
    \item \textbf{Entanglement (bipartite)}:
    \begin{itemize}
        \item \emph{Midpoint protocol}: This entanglement distribution task involves three nodes, namely two end-nodes and a central node. The two end-nodes connect to the central node via a channel. First, the two end-nodes prepare locally Bell-states, and send half of their Bell-states to the central node. The central node consequently performs a Bell-state measurement, thereby establishing a Bell-state between the end nodes. Depending on the implementation, the Bell-state measurement may use heralding to detect successful entanglement attempts.
        \item \emph{Midpoint-source protocol}: The channel configuration is as in the Midpoint protocol, but now the central node generates the Bell-state and distributes the qubits to the respective end nodes.
        \item \emph{Entanglement swapping}: A node performs a Bell-state measurement on two qubits belonging to different Bell-states, thereby creating another Bell-state of longer distance. After the measurement, the node sends the outcome to one of the nodes holding the new Bell-state to apply potential correction operations.
        \item \emph{Quantum teleportation}: This task corresponds to performing the quantum teleportation protocol on a pre-shared Bell-state \cite{BennettTele}.
        \item \emph{Entanglement purification}: Corresponds to different kinds of entanglement purification protocols \cite{Bennett96,Bennett96_2,Deutsch96,Riera2021} for Bell-states. These protocols require several input entangled states and distill entanglement into fewer copies via LOCC.
        \item \emph{Entanglement pumping}: This task corresponds to entanglement pumping protocols for Bell-states \cite{Dur2007}, similar in spirit to entanglement purification protocols. It uses two input states and pumps entanglement into one of them via LOCC.
    \end{itemize}
    \item \textbf{Entanglement (multipartite)}:
    \begin{itemize}        
        \item \emph{Entanglement purification}: Corresponds to different kinds of entanglement purification protocols for multipartite states \cite{Aschauer2005,Kruszynska2006}. These protocols require several input entangled states and distill entanglement into fewer copies via LOCC.        
        \item \emph{Graph state vertex merging}: This task merges two vertices of two graph states into a single vertex, thereby outputting another graph state, as often used in quantum network utilizing graph states \cite{Pirker_2018}. This task involves LOCC.
        \item \emph{Graph state vertex cutting}: Using this task a node can remove a vertex from a graph state by applying a Pauli Z measurement \cite{heinEntanglementGraphstates}. Depending on the measurement outcome, correction operations are necessary.
        \item \emph{Graph state local complementation}: The local complementation task performs a local complementation of a graph state by applying LOCC \cite{heinEntanglementGraphstates}.
        \item \emph{Graph state fission}: This task splits a given graph state at a vertex into two disjoint graphs, each having some of the edges preserved \cite{miguelramiro2024graph}.
        \item \emph{MBQC tasks}: Tasks of this kind utilize a shared entangled state across the network to perform a distributed quantum computation \cite{Beals20120686,CiracDistributed,CALEFFI2024110672,Cacciapuoti18,Caleffi18a}, like for example a remote gate, or a circuit via measurement-based quantum computing on cluster states \cite{Raussendorf2001,Raussendorf2003,Briegel2009} or via states corresponding to Jamiolkowski states \cite{Jamiol}. 
    \end{itemize}
    \item \textbf{Classical Messaging}:
    \begin{itemize}
        \item \emph{Send a classical message to another node}: Using this task a node sends a classical message to another node in the network.
        \item \emph{Broadcast a classical message}: This task corresponds to broadcasting a classical message in the network to all other nodes.
    \end{itemize}
    \item \textbf{Operations}:
    \begin{itemize}
        \item \emph{Apply operation}: This basic task corresponds to applying a unitary or measurement to qubits stored by a node.
        \item \emph{Prepare state}: This task corresponds to preparing a state locally within a quantum network node.
        \item \emph{QEC encoding task}: Using this task a node encodes a single qubit into several qubits using quantum error correction.
        \item \emph{QEC decoding task}: To inverse the encoding task for quantum error correction, a node applies the decoding task.
    \end{itemize}
\end{itemize}

\subsubsection{Capabilities}

When composing a new saga from an objective the capabilities of individual quantum network nodes are crucial. For example, some network nodes may not be able to store qubits due to the lack of quantum memory, which implies that many entanglement tasks such as entanglement purification or merging graph states may not be available. To take into account for the capabilities of quantum network nodes it is necessary to exchange information about which tasks a quantum network can execute and implement. 

\begin{figure}
    \includegraphics[width=\columnwidth]{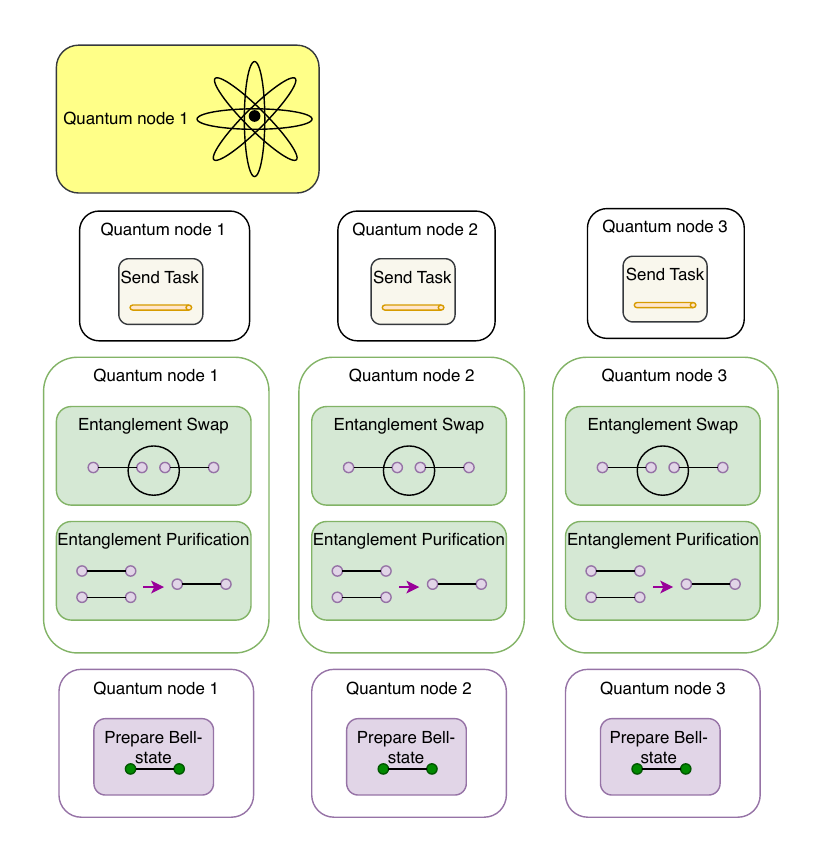}
    \caption{Quantum network node 1 knows about all the capabilities of all other nodes in the same logical network (channel- and entanglement-wise). In the figure above all nodes have the same capabilities for all resource types.}
    \label{fig:capabilities_view}
\end{figure}

The scheme we propose here, see also Fig. \ref{fig:capabilities_view}, is that all quantum network nodes have a global view on all the capabilities of each and every node in a quantum network. A capability here refers to tasks which a quantum network node can execute. We note that such tasks comprise classical messages and quantum operations, where such operations introduce noise to quantum states. Hence, not only the type of the task must be distributed but also the noise that the task introduces. Furthermore, some capabilities, i.e. feasible tasks, change over time, like tasks utilizing entanglement from the network.

To distribute capabilities, every quantum network node advertises its capabilities on start-up. By advertisement we mean here broadcasting a classical message to all other network nodes.

Capabilities are vital for the network resource manager of a quantum network node when composing a new saga. The manager takes into account all the capabilities of all quantum network nodes on such compositions and derivations.

\subsection{Sagas}

Objectives describe a certain goal of a quantum network without having a concrete way how to achieve that goal. A saga in turn achieves an objective by composing tasks from the resources of the network. One can think of a saga as a distributed workflow, comprising multiple tasks which run on quantum network nodes. The network resource manager of a quantum network node uses the topological views and capabilities of the resources of the network and selects tasks capable of achieving the desired objective. We note that such an algorithm to derive a saga is highly non-trivial, and depends on several factors. However, the composition of a saga into tasks is crucial for the performance of a network. In general, tasks within a saga may run either sequentially or in parallel.

\subsubsection{Saga composition}

The network resource manager of a node is responsible for transforming an objective into a saga. For that purpose, the manager uses the tasks from the resources, together with their  metrics such as fidelity of pre-established entangled states or noise in quantum operations and channels, to determine a saga which achieves the objective. We note that concrete implementations of such a network resource manager go far beyond the scope of this work. However, some ideas should be sketched here.

A network resource manager could for example always prioritize using pre-established entanglement before using quantum channels to distribute fresh entanglement. This reduces waiting times for clients, however, it necessitates a clever implementation for the resource monitor to replenish utilized entanglement in the network. On the other hand, a network resource manager could learn how task compositions influence the final quality of states, and make decisions based on learnt information.

\begin{figure}[h!]
    \includegraphics[width=\columnwidth]{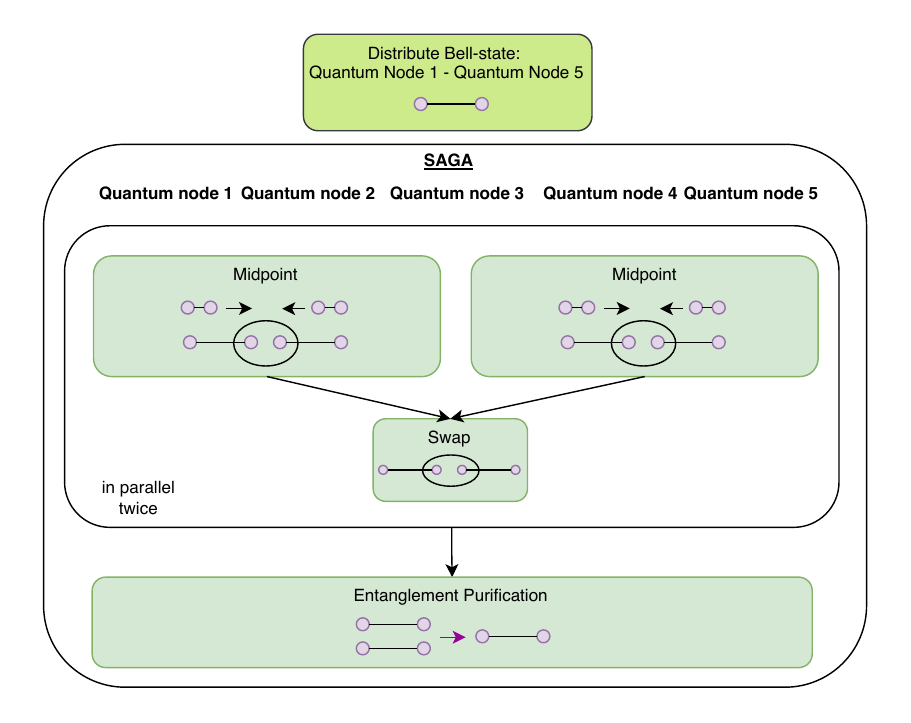}
    \caption{A saga to establish a Bell-state between node 1 and node 5. Node 1, 2 and 3 execute a Midpoint task, as also node 3, 4 and 5 do. After completing them, an entanglement swap is performed at node 3. All nodes repeat these tasks twice, and ultimately perform entanglement purification afterwards.}
    \label{fig:saga_1}
\end{figure}

\begin{figure}
    \includegraphics[width=\columnwidth]{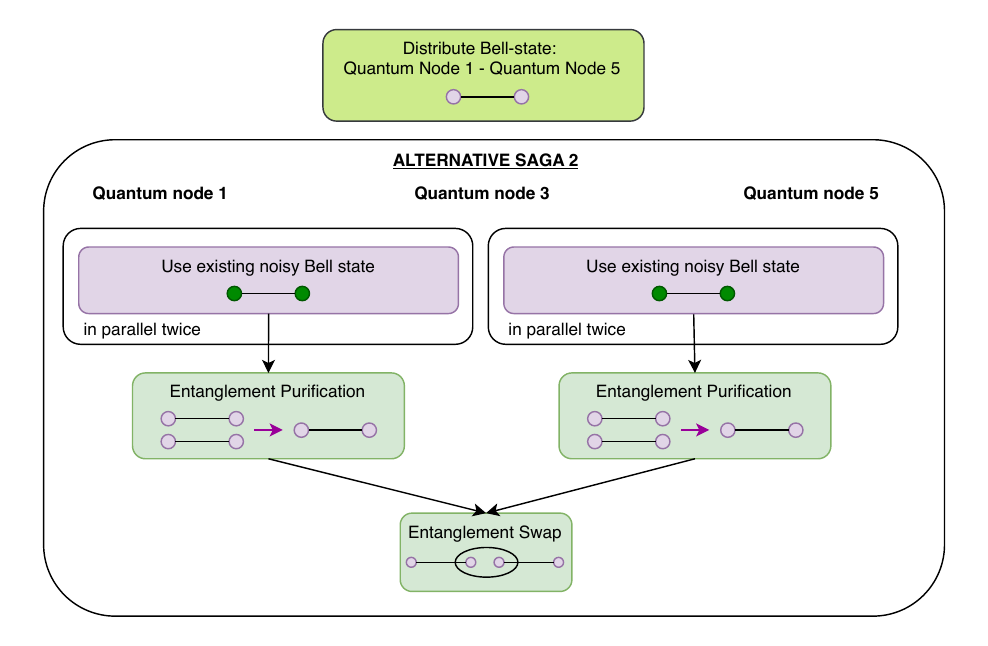}
    \caption{An alternative saga to establish a Bell-state between node 1 and node 5 which uses pre-shared entanglement between node 1 and node 5.}
    \label{fig:saga_2}
\end{figure}

We stress that the composition of a saga will heavily influence the ultimate fidelity and quality of a state. For instance, parallelizing tasks as much as possible will certainly minimize quantum memory times for qubits, and this in turn will positively impact the achievable fidelity. Also the chosen tasks itself which the saga comprises will influence the final fidelity. We point out that in order to achieve an objective, multiple sagas are valid, see Fig. \ref{fig:saga_1} and Fig. \ref{fig:saga_2} which provide two alternative sagas to distribute a Bell-state.

The saga shown in Fig. \ref{fig:saga_1} composes Midpoint tasks and entanglement swaps in parallel, together with a consequent entanglement purification step. In contrast, the saga of Fig. \ref{fig:saga_2}, uses two established Bell-state between node 1 and node 3 and node 3 and node 5 respectively to perform entanglement purification followed by a swapping step. 

We note that both sagas accomplish the same objective, namely establishing a Bell-state between node 1 and node 5. The choice of which saga is favorable in certain situations may depend on current traffic and workload of the network and its nodes, the quality of operations etc.

\subsubsection{Execution of sagas}

\begin{figure}[h!]
    \includegraphics[width=\columnwidth]{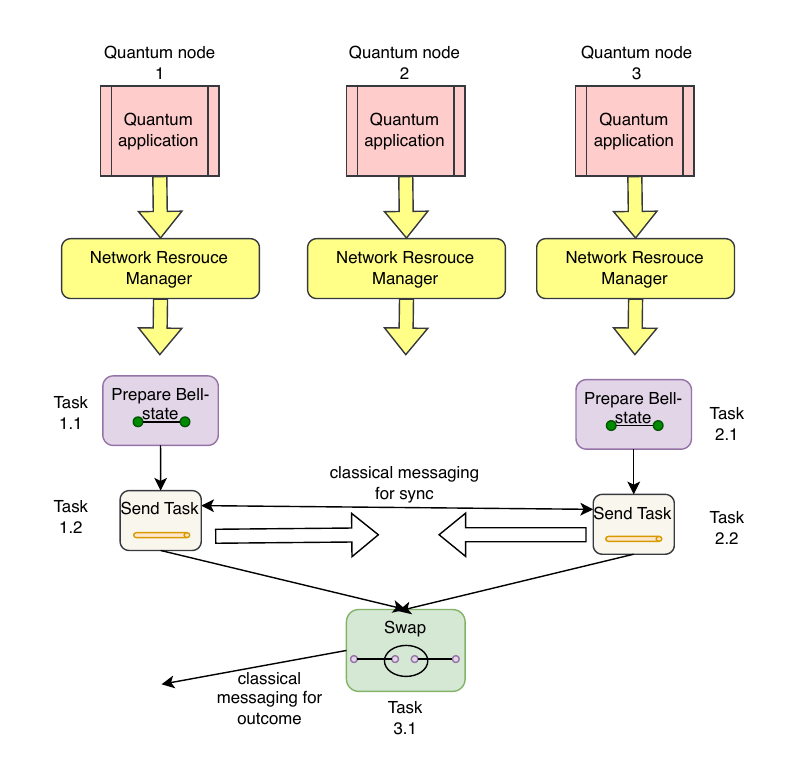}
    \caption{Execution of a saga in a quantum network: Quantum applications use the network resource manager to execute tasks. In this example, node 1 and node 3 both prepare Bell-states which they consequently send in a synchronized manner to node 2. Node 2 ultimately performs an entanglement swap task to generate a Bell-state between node 1 and node 3.}
    \label{fig:saga_exec}
\end{figure}

When the quantum network nodes execute a saga, the corresponding tasks send classical messages between the nodes to coordinate the operations of the nodes. It is also important to note that sagas itself require classical management messages to start the saga, to start tasks as part of the saga or to end a saga. The execution of a saga follows the depiction in Fig. \ref{fig:saga_exec}.

Executing a saga can either be done in orchestration, or choreography. Orchestration requires the saga-initiator to control the saga execution flow and kick-off each task that must be executed by nodes. In contrast, in saga executed by choreography the nodes notify all consecutive nodes to start the next task in a distributed manner. Both modes of operation have advantages and disadvantages, however, the framework we introduce here can handle both situations.

\begin{figure}[h!]
    \includegraphics[width=\columnwidth]{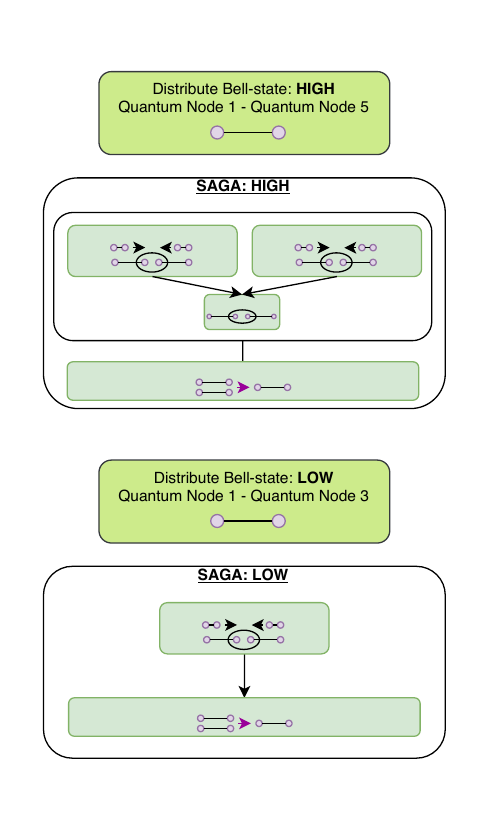}
    \caption{Objectives have priorities, and these priorities are set accordingly also to sagas to manage scheduling of tasks in other quantum network nodes.}
    \label{fig:priorities}
\end{figure}

Lastly, we also comment on the priority of sagas and objectives, see also Fig. \ref{fig:priorities}. Each objective may come with an associated priority for its completion. For example, a security application may need to establish a secret key by the means of QKD protocols urgently to deliver a classical message. The resource manager must take into account for such priorities when deriving, and especially scheduling the execution of sagas. The priority of an objective is reflected in the priority of saga, hence providing this aforementioned priority of saga execution also to other quantum network nodes participating in a saga.

\section{Outlook and conclusions}\label{sec:outlook}

In this work we proposed a resource-centric task-based quantum network control framework. We highlight that the lack of hierarchy in our model streamlines the execution of tasks, and minimizes the performance overhead of nodes. This in turn aids in reducing noise stemming from quantum memory on persisted qubits, and thereby enhances the quality of persisted states. Our model structure flattens out any hierarchy and reflects the very dynamic nature of quantum networks by categorizing the resources of a quantum network into three different types, namely channels, entanglement and classical messaging. These resources correspond to the elementary resource in a quantum network, and our approach allows to take into account for the different topologies but also capabilities of nodes in a quantum network. By separating objectives from sagas we achieved a maximum of flexibility in quantum network control. This takes into account for the diverse hardware platforms and capabilities in a quantum network, as the objective itself is independent of the composition of executable tasks in a saga.  Furthermore, the model also clearly aids standardization efforts, as fine-granular tasks may serve as units subject to standards to be composed into larger workflows via sagas. 

From this proposal several new research directions emerge. For example, a concrete implementation of the network resource manager appears vital for having a running quantum network node. The difficulty in implementing such a network resource manager lies in the complexity of saga derivation. The composition of a saga into tasks should take into account for numerous factors, such as the topologies, timings, noise in gates, noise in channels, pre-shared entanglement and workload of the network amongst other parameters. Therefore, a future possible line of research corresponds to exploring implementations of the network resource manager.

But also the performance of tasks and sagas offers various different topics. For example, it would be interesting to compare and benchmark different saga compositions for a single objective. This will shed light into the performance of a quantum network in certain situations. 

Furthermore, identifying and discussing individual, novel tasks will help completing the model. Tasks can also be easily seen as units for standardizations, as they define the classical messages together with the quantum operations necessary to achieve an objective.

\begin{acknowledgments}  
This research was funded in whole or in part by the Austrian Science Fund (FWF) 10.55776/P36010. For open access purposes, the author has applied a CC BY public copyright license to any author accepted manuscript version arising from this submission.
\end{acknowledgments}  

\bibliography{main}

\end{document}